\documentclass[preprint,showpacs,preprintnumbers,amsmath,amssymb,floatfix,nofootinbib]{revtex4}

\usepackage[breaklinks=true,colorlinks=true,backref,pagebackref]{hyperref}

\usepackage{graphicx}
\usepackage{dcolumn}
\usepackage{bm}

\usepackage{graphicx,color}

\def\a{\alpha}
\def\b{\beta}
\def\g{\gamma}
\def\d{\delta}

\def\vt{\vartheta}

\def\ba{\begin{eqnarray}}
\def\ea{\end{eqnarray}}
\def\w{\wedge}

\begin{document}

\title{Non-minimally coupled Dirac equation with torsion:\\
Poincar\'e gauge theory of gravity with even and odd parity terms}

\author{ Muzaffer Adak  \\
 {\small Department of Physics, Faculty of Arts and Sciences,} \\
 {\small Pamukkale University, 20017 Denizli, Turkey} \\
 {\small \tt madak@pau.edu.tr}
  }

\date{{13 March 2012}, {\it file DiracTorsion11g.tex}}

 \begin{abstract}
  We take a Dirac field non-minimally coupled to the gravitational field
  within the framework of the Poincar\'e gauge theory of gravity with
  torsion and curvature. We study the subcase of ``weak'' gravity,
  that is, the gravitational Lagrangian depends only linearly on the
  curvature and quadratically on the torsion. We include all pieces in
  curvature and torsion that are of odd parity. The second
  field equation of gravity is derived by varying the Lorentz
  connection. We solve it with respect to the torsion and decompose the first
  field equation of gravity and the Dirac equation into Einsteinian pieces and
  post-Riemannian terms.
 \end{abstract}

\pacs{04.50.Kd, 03.65.Pm, 11.15.-q} 

\maketitle

\section{Introduction}
When parity violation in quantum gravity is addressed, one
generally studies the coupling of fermionic degrees of freedom in
the presence of torsion. This is done by reading out the most
relevant terms in the low-energy effective lagrangian. In the
Poincare gauge theory of gravity the dynamical variables are given
by orthonormal coframe $\vt^\a$ and the Lorentz connection
$\Gamma^{\a\b}$. If one applies the effective field theory point
of view to gravity in the low-energy limit, then to leading order
the low-energy effective lagrangian contains exactly six terms,
see e.g. \cite{Freidel2005} and references therein. These are the
Euler term $R^{\a\b} \w R^{\g\d} \epsilon_{\a\b\g\d}$, the
Pontryagin term $R^{\a\b} \w R_{\a\b}$, the Nieh-Yan term $T_\a \w
T^\a + R_{\a\b} \w \vt^{\a\b}$, the Einstein-Hilbert term
$R_{\a\b} \w \eta^{\a\b}$, the cosmological term $\lambda_0 \eta$
and the curvature pseudoscalar $R_{\a\b} \w \vt^{\a\b}$. Since the
first two of them are topological invariants and the Nieh-Yan term
is an exact form, people concentrate on the last three bulk terms
as {\it the most} general four-dimensional low-energy
gravitational lagrangian. When a Dirac field is minimally coupled
to that gravitational lagrangian with its three bulk terms, the
parity-violating vector-axial interaction is absent and
correspondingly parity is not violated by the gravitational field.
If the coupling is non-minimal, then one can see that the
parity-violating term is proportional to the Immirzi parameter
\cite{Freidel2005}, \cite{Khriplovich2005}. That {\it most
general} gravitational lagrangian corresponds to the first three
terms of our lagrangian (\ref{weakadak}). We compare with the
literature and find that the coupling coefficient $b_0$ is
proportional to the inverse of the Immirzi parameter.

The parity preserving $R+R^2+T^2$-type theory (with or without
matter fields) have been studied in the early 1980s by different
groups for different aims in the literature. For example, in
Ref\cite{Hayashi1980} Hayashi and Shirafuji minimally coupled a
Dirac field and analyzed the field equations, in
Ref\cite{Blagojevic1983} Blagojevic and Nikolic investigated
general aspects of the Hamiltonian structure of the Poincare gauge
theory of gravity of that type in the time gauge and in
Ref\cite{Nikolic1984} Nikolic applied Dirac's Hamiltonian method
for constrained systems to that type of gravity interacting with
an arbitrary matter field. In a recent paper \cite{Baekler:2010fr}
Baekler et al proposed a cosmological model in the framework of
the Poincare gauge theory of gravity of that type with even parity
and odd parity.

In general, there are two types of actions: minimal and
non-minimal. The non-minimal interactions with spinors and scalars
provide certain advantages at the quantum level, for they give the
possibility to construct renormalizable theory. For more detailed
discussion on non-minimal couplings see the review
\cite{Shapiro2002}. Our wish here is to investigate the possible
parity effects of a non-minimally coupled Dirac field to {\it the
most general} ``weak'' gravitational interactions by studying the
most general $R+T^2$-type gravity lagrangian including all
possible even and odd parity terms. As it contains eight
parameters $\alpha,\lambda_0,a_0,b_0,k_1,k_2,k_3,\ell_2$ in
general, the non-minimally Dirac coupled system can be described
by at most seven parameters ($k_1=0$ without loss of generality).

What we do is basically to consider that total lagrangian and to
derive the equations of motions via independent variations in
accord with the orthonormal coframe, the Lorentz connection
1-forms and the Dirac spinor. Then we solve explicitly the SECOND
field equation for the torsion tensor. At the next step we rewrite
the remaining field equations naively in terms of the familiar
Riemannian terms plus other terms depending on the Dirac field and
its current vectors and pseudo-vectors. Finally we give the
equivalent lagrangian written in the Riemannian spacetime. This
form of the theory should also be useful in view of future
experiments that could put constraints on the parameters of the
theory, and test possible effects resulting from a non-vanishing
torsion. The same strategy was followed in \cite{Purcell1978} in
which the gravity lagrangian contains only three parameters
corresponding to our coupling coefficients $a_0, b_0, \ell_2$ and
the matter coupling is minimal. Correspondingly the {\it novelty}
of our work is to display the most general gravity model
containing the non-minimal coupled Dirac field and the
non-propagating torsion including all eight parameters known from
the literature.

\section{Mathematical Preliminaries} \label{sec:MathPreli}

Spacetime is denoted by the triple $\{M,g,\nabla \}$ where $M$ is
differentiable 4-dimensional manifold, $g=g_{ij}(x) dx^i \otimes
dx^j = g_{\alpha \beta} \vartheta^\alpha (x) \otimes
\vartheta^\beta (x)$ is the metric tensor with $g_{\alpha \beta} =
diag(-1,1,1,1)$, and $\nabla$ represents the linear connection. We
require that $M$ is orientable, otherwise one has problems with
defining the volume form and the Hodge dual. We take the
conventions that $\alpha , \beta , \cdots = 0,1,2,3$ denote the
orthonormal (anholonomic) indices and $i,j, \cdots = \hat{0},
\hat{1}, \hat{2}, \hat{3}$ the coordinate indices.
Correspondingly, $\vartheta^\alpha (x) = e_i{}^\alpha(x) dx^i$,
where $dx^i$ is a coordinate coframe and $e_i{}^\alpha (x)$ are
tetrad components (or coordinate components of orthonormal
coframe). Thus as $ e_\alpha(x)$ denotes the orthonormal frame,
$\partial_i$ the coordinate frame such that $e_\alpha(x) =
e_\alpha{}^i(x) \partial_i$. Frame and coframe are dual to each
other,
 \begin{eqnarray}
          \vartheta^\alpha(e_\beta)\equiv
          e_\beta \lrcorner \vartheta^\alpha=\delta^\alpha_\beta \, ,
 \end{eqnarray}
where $ \delta^\alpha_\beta$ is the Kronecker symbol and
$\lrcorner$ denotes the interior product. The linear connection is
determined by local (linear) connection 1-forms
$\Gamma_\a{}^\b(x)$ as follows: $\nabla_{e_\a}e_\b = -
\Gamma_\b{}^\g(e_\a) e_\g$. We prefer to work with orthonormal
frame because of its coordinate independence. It is also necessary
when one wants to introduce spinors. With the shorthand notation
$\vartheta^\alpha \wedge \vartheta^\beta \wedge \cdots \equiv
\vartheta^{\alpha \beta \cdots }$ and the volume 4-form $\eta :=
{}^\star1$ we use the following eta-basis,
 \ba
    \eta^\alpha &:=& e^\alpha \lrcorner \eta = {}^\star\vartheta^\alpha \, ,
    \nonumber \\
    \eta^{\alpha \beta} &:=& e^\beta \lrcorner \eta^\alpha =
    {}^\star\vartheta^{\alpha \beta} \, , \nonumber \\
    \eta^{\alpha \beta \gamma} &:=& e^\gamma \lrcorner \eta^{\alpha \beta} =
    {}^\star\vartheta^{\alpha \beta \gamma}  \, , \nonumber \\
    \eta^{\alpha \beta \gamma \rho} &:=& e^\rho \lrcorner \eta^{\alpha \beta \gamma} =
    {}^\star\vartheta^{\alpha \beta \gamma \rho} \, ,
 \ea
where $\star$ denotes Hodge star.

A geometry (in orthonormal frames) is determined by the Cartan structure equations
 \ba
     Q_{\alpha \beta} &:=& - D g_{\alpha \beta}
                      =  \Gamma_{\alpha \beta} + \Gamma_{\beta \alpha} \, , \label{nonmet}\\
   T^\alpha &:=& D \vartheta^\alpha = d \vartheta^\alpha +\Gamma_\beta{}^\alpha \wedge \vartheta^\beta \, , \label{tors}\\
     R_\alpha{}^\beta &:=& D \Gamma_\alpha{}^\beta := d \Gamma_\alpha{}^\beta
                  - \Gamma_\alpha{}^\gamma \wedge \Gamma_\gamma{}^\beta \, ,\label{curva}
 \ea
where $Q_{\alpha \beta}$ are the nonmetricity 1-forms, $T^\alpha$
the torsion 2-forms and $ R_\alpha{}^\beta$ the curvature 2-forms.
The operators $d$ and $D$ denote the exterior derivative and the
covariant exterior derivative, respectively. These tensors satisfy
the Bianchi identities
 \ba
       D Q_{\alpha \beta} &=&   R_{\alpha \beta} +R_{\beta \alpha} \; , \label{bianc:0} \\
       D T^\alpha    &=& R_\beta{}^\alpha \wedge \vartheta^\beta \; , \label{bianc:1} \\
       D R_\alpha{}^\beta &=& 0  \; . \label{bianc:2}
 \ea
Differentiating the $\eta$'s, we find the following useful relations
 \ba
    D \eta_\alpha &=& -2Q \wedge \eta_\alpha + T^\epsilon \wedge \eta_{\alpha \epsilon}\, , \nonumber \\
    D \eta_{\alpha \beta} &=& -2Q \wedge \eta_{\alpha \beta} + T^\epsilon \wedge \eta_{\alpha \beta \epsilon}\, , \nonumber \\
    D \eta_{\alpha \beta \gamma} &=& -2Q \wedge \eta_{\alpha \beta \gamma} + T^\epsilon \wedge \eta_{\alpha \beta \gamma \epsilon}\, , \nonumber \\
    D \eta_{\alpha \beta \gamma \delta} &=& -2Q \wedge \eta_{\alpha \beta \gamma \delta} \, ,
 \ea
where $Q=Q^\alpha{}_\alpha /4$ is the Weyl covector.

In this work we set nonmetricity to zero. This means, in orthonormal frames,
$\Gamma_{\alpha \beta} = - \Gamma_{\beta \alpha}$ and $R_{\alpha
\beta} = -R_{\beta \alpha}$. That spacetime is known as the
Riemann-Cartan spacetime with metric compatible connection. In
this case, the linear connection 1-forms can be decomposed as
follows \cite{Hehl:1994ue},\cite{Dereli1996}:
 \ba
     \Gamma_\alpha{}^\beta = \widetilde{\Gamma}_\alpha{}^\beta - K_\alpha{}^\beta \, , \label{connect:dec}
 \ea
where the contortion 1-forms $K_\alpha{}^\beta \w \vt_\b = T_\a$ or
 \ba
   2 K_{\alpha \beta} = e_\alpha \lrcorner T_\beta -e_\beta \lrcorner T_\alpha
           - (e_\alpha \lrcorner e_\beta \lrcorner T_\gamma) \vartheta^\gamma  \, , \label{contort}
 \ea
the Christoffel 1-forms (or Riemannian connection 1-forms)
$\widetilde{\Gamma}_\alpha{}^\beta$
 \ba
     2 \widetilde{\Gamma}_{\alpha \beta} = d g_{\alpha \beta} + (e_\alpha \lrcorner dg_{\beta \gamma} - e_\beta \lrcorner dg_{\alpha \gamma}) \vartheta^\gamma + e_\alpha \lrcorner d\vartheta_\beta -e_\beta \lrcorner d\vartheta_\alpha
           - (e_\alpha \lrcorner e_\beta \lrcorner d\vartheta_\gamma) \vartheta^\gamma \, . \nonumber
 \ea
In orthonormal frames $\widetilde{\Gamma}_{\alpha \beta}$ turns out to be
 \ba
     2 \widetilde{\Gamma}_{\alpha \beta} = e_\alpha \lrcorner d\vartheta_\beta -e_\beta \lrcorner d\vartheta_\alpha
           - (e_\alpha \lrcorner e_\beta \lrcorner d\vartheta_\gamma) \vartheta^\gamma \, . \label{christoffel}
 \ea
In this work riemannian quantities are labeled by a tilde, e.g.
Riemann (curvature) 2-form
 \ba
    \widetilde{R}_\alpha{}^\beta = d \widetilde{\Gamma}_\alpha{}^\beta
                  - \widetilde{\Gamma}_\alpha{}^\gamma \wedge
                  \widetilde{\Gamma}_\gamma{}^\beta \; .
                  \label{riemanncurv}
 \ea

\section{Gravitational Lagrangian}

In the framework of the Poincar\'e gauge theory, the gravitational
lagrangian is composed of parity even and parity odd pieces \cite{Baekler:2010fr}:
\begin{equation}\label{LagrangianPM}
V_{\pm}=V_{+}+V_{-}\,,
\end{equation}
with
\begin{eqnarray}
  V_{+}  = \frac{1}{2\kappa}\left(-a_0 \,^{(6)}R^{\alpha\beta}\wedge\eta_{\alpha\beta} -2\lambda_{0}\eta
           + T^\alpha\wedge\textstyle\sum\limits_{I=1}^{3}a_{I}{}^{\star(I)} T_\alpha \right)
           - \frac{1}{2\varrho} \left(R^{\alpha\beta} \wedge{}\textstyle\sum\limits_{I=1}^{6} w_{I}{}^{\star(I)} R_{\alpha\beta}\right) \label{parityeven}
\end{eqnarray}
and
\begin{eqnarray}
 V_{-}&=& -\frac{b_0}{2\kappa}\,^{(3)}R_{\alpha\beta} \wedge\vt^{\a\b}
       + \frac{1}{\kappa}\left( {\sigma}_{1}{}^{(1)}T^{\alpha}\wedge\, ^{(1)}T_{\alpha}
       + {\sigma}_{2}{}^{(2)}T^{\alpha}\wedge{} ^{(3)}T_{\alpha}\right) \label{parityodd} \\
    & &-\frac{1}{2{\varrho}}\left( {\mu}_{1}{}^{(1)}R^{\alpha\beta}\wedge{}^{(1)}R_{\alpha\beta}
       + {\mu}_{2}{}^{(2)}R^{\alpha\beta}\wedge{}^{(4)}R_{\alpha\beta}
       + {\mu}_{3}{}^{(3)}R^{\alpha\beta}\wedge{} ^{(6)}R_{\alpha\beta}
       + {\mu}_{4}{}^{(5)}R^{\alpha\beta}\wedge{} ^{(5)}R_{\alpha\beta} \right) \nonumber
\end{eqnarray}
where $\kappa$ and $\varrho$ are respectively the weak and strong
gravitational constants, $a_0,b_0,a_I,w_I,\sigma_I,\mu_I$ are the
dimensionless coupling constants and $\lambda_0$ is the
cosmological constant, ${}^{(I)}T^{\alpha}$ are the irreducible
decompositions of the torsion and ${}^{(I)}R_{\alpha\beta}$ are
the irreducible pieces of the curvature. This lagrangian obtained
from a classic field theoretical view point is equivalent to one
\cite{Diakonov2011} obtained from a quantum field theoretical view
point, see \cite{Baekler2011}. If we specialize to the subcase of
{\it weak} gravity, that is, we drop the terms that are multiplied
by $1/\varrho$, then the lagrangian reads
\begin{eqnarray}\nonumber
\label{weak}
V_{\pm}|_{\text{weak}}&=&
  \frac{1}{2\kappa}(-a_0\,^{(6)}R^{\alpha\beta}\wedge\eta_{\alpha\beta}
    -2\lambda_{0}\eta-b_0 \,^{(3)}R_{\alpha\beta}
  \wedge\vt^{\a\b}\\  &&
    \hspace{20pt} +T^\alpha\wedge\textstyle\sum
    \limits_{I=1}^{3}a_{I}{}^{\star(I)}T_\alpha+ 2 {\sigma}_{1}{}
^{(1)}T^{\alpha}\wedge\, ^{(1)}T_{\alpha} + 2 {\sigma}_{2}{}
^{(2)}T^{\alpha}\wedge{} ^{(3)}T_{\alpha})\,.\nonumber
\end{eqnarray}
By noticing the trace and symmetry properties of the irreducible
decompositions of curvature $\,^{(3)}R_{\alpha\beta}
\wedge\vt^{\a\b}=R_{\alpha\beta} \wedge\vt^{\a\b} $ and
$\,^{(6)}R^{\alpha\beta}\wedge\eta_{\alpha\beta} =
R^{\alpha\beta}\wedge\eta_{\alpha\beta}$, more explicitly, we have
\begin{eqnarray}
     V_{\pm}|_{\text{weak}} &=& \frac{1}{2\kappa}(-a_0R^{\alpha\beta}\wedge\eta_{\alpha\beta}
    -2\lambda_{0}\eta-b_0 R_{\alpha\beta} \wedge\vt^{\a\b} + a_{1}\,^{(1)}T^\alpha\wedge{}^{\star(1)}T_\alpha \label{weakhehl} \\
& & + a_{2}\,^{(2)}T^\alpha\wedge{}^{\star(2)}T_\alpha  +a_{3}\,^{(3)}T^\alpha\wedge {}^{\star(3)}T_\alpha
  + 2 {\sigma}_{1}{}^{(1)}T^{\alpha}\wedge\, ^{(1)}T_{\alpha} + 2 {\sigma}_{2}{}^{(2)}T^{\alpha}\wedge{} ^{(3)}T_{\alpha})\,.\nonumber
\end{eqnarray}
This lagrangian 4-form is equivalent to
 \begin{eqnarray}
    V_{\pm}|_{\text{weak}} &=& \frac{1}{2\kappa}(-a_0R^{\alpha\beta}\wedge\eta_{\alpha\beta}
    -2\lambda_{0}\eta-b_0 R_{\alpha\beta} \wedge\vt^{\a\b} + k_1 T^\alpha \w {}^\star T_\alpha \label{weakadak} \\
    & &+ k_2 (e_\alpha \lrcorner T^\alpha) \w {}^\star (e_\beta \lrcorner T^\beta)
    + k_3 (\vt_\alpha \w T^\alpha) \w {}^\star (\vt_\beta \w T^\beta)\nonumber \\
  & & + 2 \ell_1 T^\a \w T_\a + 2\ell_2  (e_\alpha \lrcorner T^\alpha) \w (\vt_\beta \w T^\beta))\,.\nonumber
\end{eqnarray}
under the redefinition of the coupling coefficients
 \ba
   a_1 = k_1 \, , \; a_2=k_1+3k_2 \, , \; a_3 = k_1 + 3k_3 \, , \; \sigma_1 = \ell_1 \, , \; \sigma_2=2\ell_1-3\ell_2 \, .
 \ea
This correspondence was checked by Reduce-Excalc
\cite{Hearn1093},\cite{Schrufer}. The term with $b_0$ or
$\sigma_1$ in (\ref{weakhehl}) or, equivalently, that with $b_0$
or $\ell_1$ in (\ref{weakadak}) can be dropped by using the parity
odd boundary term, the so-called Nieh-Yan 4-form
\cite{Nieh1982},\cite{Nieh2007},
$B_{TT}^-=dC_{TT}^-=\frac{1}{2}d(\vt^\a \w T_\a)=\frac{1}{2}(T_\a
\w T^\a + R_{\a\b} \w \vt^{\a\b})$. We will choose
$\sigma_1=\ell_1=0$ without loss of generality.

\subsection*{Dirac Lagrangian}

We are using the formalism of Clifford algebra
$\mathcal{C}\ell_{1,3}$-valued exterior forms. The
$\mathcal{C}\ell_{1,3}$ algebra is generated by the relation among
the orthonormal basis $\{\gamma_0,\gamma_1,\gamma_2,\gamma_3\}$
 \begin{equation}
    \gamma^\alpha \gamma^\beta + \gamma^\beta \gamma^\alpha = 2g^{\alpha \beta} \, .
 \end{equation}
One particular representation of the $\gamma^\alpha$'s is given by the following Dirac matrices
 \begin{eqnarray}
   \gamma_0 = i\left(\begin{array}{cc}
                -I & 0 \\
                0 & I
              \end{array}\right) \, , \;
   \gamma_1 = i\left(\begin{array}{cc}
                0 & \sigma^1 \\
                -\sigma^1 & 0
              \end{array}\right) \, , \;
   \gamma_2 = i\left(\begin{array}{cc}
                0 & \sigma^2 \\
                -\sigma^2 & 0
              \end{array}\right) \, , \;
  \gamma_3 = i\left(\begin{array}{cc}
                0 & \sigma^3 \\
                -\sigma^3 & 0
              \end{array}\right) \, ,
 \end{eqnarray}
where $\sigma^1,\sigma^2,\sigma^3$ are the Pauli matrices. In this
case a Dirac spinor $\Psi$ can be represented by a 4-component
column matrix. Thus we write explicitly the covariant exterior
derivative of $\Psi$ and the quantity $D^2\Psi$
 \ba
  D\Psi = d\Psi - \frac{1}{2} \Gamma^{\alpha \beta} \sigma_{\alpha \beta} \Psi \; , \quad
  D^2 \Psi = - \frac{1}{2} R^{\a\b} \sigma_{\alpha \beta} \Psi \, ,
 \ea
where $\sigma_{\alpha \beta}:= \frac{1}{4}[\gamma_\alpha ,
\gamma_\beta]$ are the generators of the Lorentz group
\cite{Adak2003},\cite{Adak2010}. The Dirac adjoint is
$\overline{\Psi}:= \Psi^\dag \gamma_0$.  Some relations of the
Dirac matrices are
 \ba
 \begin{array}{rcl}
   \sigma_{\alpha \beta}\gamma_\gamma & =& \gamma_\gamma \sigma_{\alpha \beta}+ g_{\beta \gamma} \gamma_\alpha - g_{\alpha \gamma} \gamma_\beta \,,\\
   \sigma_{\alpha \beta} \gamma_\gamma + \gamma_\gamma \sigma_{\alpha \beta} &=& - \epsilon_{\alpha \beta \gamma \delta} \gamma^\delta \gamma_5 \,,\\
   \gamma_\g \sigma_{\a\b} &=& \frac{1}{2} g_{\a\g} \gamma_\b - \frac{1}{2} g_{\b\g} \gamma_\a - \frac{1}{2}\epsilon_{\a\b\g\delta} \gamma^\delta \gamma_5 \, , \\
   {}[\sigma_{\alpha \beta}, \sigma_{\gamma \delta}] &=& -g_{\alpha \gamma} \sigma_{\beta \delta} - g_{\beta \delta} \sigma_{\alpha \gamma} + g_{\alpha \delta} \sigma_{\beta \gamma} + g_{\beta \gamma} \sigma_{\alpha \delta} \, ,
 \end{array} \label{dirmatrixprop}
 \ea
where $\gamma_5 := \gamma_0 \gamma_1 \gamma_2 \gamma_3$. In order
to obtain the Bjorken-Drell conventions \cite{Bjorken1964} one has
to replace $\gamma^\a \rightarrow -i \gamma^\a$ and $\gamma_5
\rightarrow i \gamma_5$. Now we take the non-minimally coupled
Dirac lagrangian given by the {\it hermitian} 4-form
\cite{Birkbook}, \cite{Adak:2001pq}
 \begin{equation}
 L_{\rm D}(g_{\a\b},\vartheta^\alpha,\Psi,D\Psi)
          = \frac{i\hbar}{2}\left[(1-i\a)\overline{\Psi}\;{}^\star\gamma\wedge D\Psi + (1+i\a)D\overline{\Psi}\wedge{}^\star\gamma\,\Psi\right]
            + imc \,\overline{\Psi}\Psi \, \eta \, , \label{10-4.10}
 \end{equation}
where $\gamma:=\gamma_\alpha \vartheta^\alpha$ and $\a$ is a real
constant. The value $\a=0$ corresponds to the conventional minimal
coupling of fermions to gravity. In general, an arbitrary real
value for $\a$ corresponds to a non-minimal coupling. The coframe
$\vartheta^\alpha$ necessarily occurs in the Dirac Lagrangian,
even in special relativity. The hermiticity of the lagrangian
(\ref{10-4.10}) leads to a charge current which admits the usual
probabilistic interpretation.

\section{Total Lagrangian and Field Equations}

In this work we will consider the total lagrangian
\begin{equation}
  L_{\text{tot}}=V_{\pm}|_{\text{weak}}(g_{\a\b},\vt^\a,T^\a,R^{\a\b})+L_{\rm D}(g_{\a\b},\vt^\a,\Psi, D\Psi) \,. \label{lagrangetotal}
\end{equation}
Then from (\ref{weakadak}) we compute the translation and the Lorentz excitations, respectively,
 \ba
    H_\a &:=& -\frac{\partial V}{\partial T^\a} = -\frac{1}{\kappa} (k_1 {}^\star T_\a + k_2 e_\a \lrcorner {}^\star \mathcal{V} + k_3 \vt_\a \w \mathcal{A} +  \ell_2 e_\a \lrcorner {}^\star \mathcal{A} - \ell_2 \vt_\a \w \mathcal{V}) \, , \label{eq:Halp}\\
    H_{\a\b} &:=& -\frac{\partial V}{\partial R^{\a\b}}  = \frac{1}{2\kappa} (a_0 \eta_{\a\b} + b_0 \vt_{\a\b}) \, , \label{eq:Halpbet}
 \ea
where $\mathcal{V}= e_\a \lrcorner T^\a$ and $\mathcal{A}=
{}^\star(\vt_\a \w T^\a)$. Then, after calculating from
(\ref{weakadak}), (\ref{eq:Halp}) and (\ref{eq:Halpbet}) the gauge
currents of energy-momentum and spin
 \ba
    E_\alpha &:=& e_\alpha \lrcorner V + (e_\alpha \lrcorner T^\beta) \wedge H_\beta + (e_\alpha \lrcorner R^{\beta \gamma}) \wedge H_{\beta \gamma} \, , \\
    E_{\alpha \beta} &:=& -\frac{1}{2} (\vartheta_\alpha \wedge H_\beta - \vartheta_\beta \wedge H_\alpha) \, ,
 \ea
respectively, we can write the FIRST field equation $D H_\alpha - E_\alpha = \Sigma_\alpha$ as
 \ba
    \frac{a_0}{2} R^{\b\g} \w \eta_{\a\b\g} +\lambda_0 \eta_\a +b_0R_{\a\b} \w \vt^\b & &\nonumber \\
    - k_1 D{}^\star T_\a - \frac{k_1}{2} [e_\a \lrcorner (T_\b \w \, {}^\star T^\b) - 2 (e_\a \lrcorner T^\b)\w \, {}^\star T_\b] & &\nonumber \\
   -k_2 D(e_\a \lrcorner {}^\star \mathcal{V}) - \frac{k_2}{2}[e_\a \lrcorner ( \mathcal{V}\w \, {}^\star \mathcal{V}) - 2 (e_\a \lrcorner T^\b)\w (e_\b \lrcorner \, {}^\star \mathcal{V})] & &\nonumber \\
   -k_3 D(\vt_\a \w \mathcal{A}) + \frac{k_3}{2}[e_\a \lrcorner ( \mathcal{A} \w \, {}^\star \mathcal{A}) + 2 (e_\a \lrcorner T^\b)\w (\vt_\b \w \mathcal{A})] & &\nonumber \\
   -\ell_2 D(e_\a \lrcorner \, {}^\star \mathcal{A} - \vt_\a \w \mathcal{V}) - \ell_2 [e_\a \lrcorner ( \mathcal{V} \w \, {}^\star \mathcal{A})
   - (e_\a \lrcorner T^\b)\w (e_\b \lrcorner \, {}^\star \mathcal{A} - \vt_\b \w \mathcal{V})] &=& \kappa \Sigma_\a \, , \label{firsteqn}
 \ea
and the SECOND field equation $D H_{\alpha \beta} - E_{\alpha \beta} = \tau_{\alpha \beta}$ as
 \ba
    \frac{a_0}{2} T^\g \w \eta_{\a\b\g}  - \frac{b_0}{2} (\vt_\a \w T_\b - \vt_\b \w T_\a)
   -\frac{k_1}{2} (\vt_\a \w \, {}^\star T_\b - \vt_\b \w \, {}^\star  T_\a) & & \nonumber \\
    - \frac{k_2}{2} [\vt_\a \w (e_\b \lrcorner {}^\star \mathcal{V}) - \vt_\b \w (e_\a \lrcorner {}^\star \mathcal{V})]
   -k_3 \vt_{\a\b} \w \mathcal{A} & &\nonumber \\
    -\frac{\ell_2}{2} [\vt_\a \w (e_\b \lrcorner {}^\star \mathcal{A})- \vt_\b \w (e_\a \lrcorner {}^\star \mathcal{A}) - 2\vt_{\a\b}\w \mathcal{V}] &=& \kappa \tau_{\a\b} \, , \label{secondeqn}
 \ea
where the Dirac energy-momentum current and the Dirac spin current are, respectively,
 \ba
\Sigma_\alpha &=& \frac{i\hbar}{2}\left[(1-i\a)\overline{\Psi}\;{}^\star(\gamma \w \vartheta_\alpha) \wedge D\Psi - (1+i\a)D\overline{\Psi}\wedge{}^\star(\gamma \w \vartheta_\alpha) \,\Psi\right] + imc \,\overline{\Psi}\Psi \, \eta_\alpha \, , \label{direnmomcurrent}\\
\tau_{\alpha \beta} &=& \frac{i\hbar}{4} [\vartheta_{\alpha \beta} \w \overline{\Psi} \gamma \gamma_5 \Psi
+ i \a \overline{\Psi} (\gamma_\a \eta_\b - \gamma_\b \eta_\a) \Psi] \, . \label{diracpscurrent}
 \ea
Here it would be useful to remark that the FIRST field equation is
obtained by varying the total lagrangian 4-form
(\ref{lagrangetotal}) with respect to $\vt^\a$ and the SECOND
field equation with respect to $\Gamma^{\a\b}$
\cite{Baekler:2010fr},\cite{Hehl:1994ue}. Finally, the variation
of (\ref{lagrangetotal}) with respect to $\overline{\Psi}$ yields
the Dirac equation
 \ba
    {}^\star \gamma \w (D - \frac{1+i\a}{2}\mathcal{V})\Psi + \frac{mc}{\hbar} \Psi \, \eta=0 \, .\label{diraceqn}
 \ea
This equation can be decomposed into a Riemannian part plus a
torsional part by simple algebra. Firstly, by noting $\mathcal{V}
= K^{\a\b}{}_\a \vt_\b$, we rewrite it as
  \ba
    {}^\star \gamma \w \widetilde{D}\Psi + \frac{mc}{\hbar} \Psi \, \eta
    - \frac{1}{2} \left( K^{\a\b\mu}\gamma_\mu \sigma_{\a\b} - (1+i\a)  K^{\a\b}{}_\a \gamma_\b \right) \Psi \, \eta=0 \, , \nonumber
 \ea
where $K_{\a\b} = K_{\a\b \g} \vt^\g$. After using the third relation of (\ref{dirmatrixprop}), we obtain
  \ba
    {}^\star \gamma \w \widetilde{D}\Psi + \left( \frac{mc}{\hbar}
     -\frac{1}{4} \mathcal{A}_\a \gamma^\a \gamma_5 + \frac{i\a}{2} \mathcal{V}_\a  \gamma^\a \right)\Psi \, \eta=0 \, , \label{diraceqn2}
 \ea
with $e_\a \lrcorner \mathcal{V} := \mathcal{V}_\a =
K_{\b\a}{}^\b$ and $ e_\a \lrcorner \mathcal{A} := \mathcal{A}_\a
= \epsilon_{\a\b\mu\nu} K^{\b\mu \nu}$. This means that in the
minimally coupled Dirac equation only the totally antisymmetric
axial component of the torsion survives. Now, we want to reexpress
(\ref{direnmomcurrent}) in a different way by means of
(\ref{diraceqn}) and its adjoint, $(D -
\frac{1-i\a}{2}\mathcal{V})\overline{\Psi} \w \, {}^\star \gamma +
\frac{mc}{\hbar} \overline{\Psi} \, \eta=0$,
 \ba
     \Sigma_\a =\frac{i\hbar}{2} \left[(1-i\a) \overline{\Psi} \gamma_\b (D_\a\Psi) - (1+i\a) (D_\a \overline{\Psi}) \gamma_\b \Psi\right]\eta^\b
 \ea
where $D_\a := e_\a \lrcorner D$. Thus we can display the
canonical energy-momentum tensor, $\Sigma_\a := \mathcal{T}_{\a\b}
\eta^\b$, for a Dirac field as $\mathcal{T}_{\a\b} =
\frac{i\hbar}{2} \left[ (1-i\a) \overline{\Psi} \gamma_\b
(D_\a\Psi) - (1+i\a)(D_\a \overline{\Psi}) \gamma_\b \Psi\right]$.
This version of the Dirac energy-momentum may be more familiar
from the literature \cite{Hehl:1997ep},  \cite{Kirsch:2001gt},
\cite{RMP}.

The next step is to deal with the SECOND equation
(\ref{secondeqn}). But, before that, we will review the symmetry
properties of the spin current 3-form. Only the first part of the
Dirac spin current 3-form vanishes, i.e. ${}^{(1)}\tau_{\a\b}=0$,
but ${}^{(2)}\tau_{\a\b}\neq 0$ and ${}^{(3)}\tau_{\a\b}\neq 0$ ,
see the appendix for details. Then, by denoting the Dirac spin
current pseudovector and vector, respectively, as
 \ba
  S^\a := i \overline{\Psi} \gamma^\a \gamma_5 \Psi \; , \quad W^\a := i \overline{\Psi} \gamma^\a \Psi
 \ea
we can calculate $T^\a$ from the SECOND equation (\ref{secondeqn})
 \ba
   T^\a = 2\kappa \left[ \vt^\a \w (c_2S + d_2 W) + e^\a \lrcorner {}^\star (c_3S + d_3 W) \right] \, , \label{eq:tors}
 \ea
where $S = S^\a \vt_\a$, $W=W^\a \vt_\a$,
 \ba
   & & c_2 = -\hbar(b_0 + 3\ell_2)/{\mathcal{D}} \, , \quad \quad d_2 = i\hbar \a (a_0 -2 k_1 - 6k_3)/{\mathcal{D}} \, , \nonumber \\
   & & c_3 = \hbar (2a_0 - k_1 + 3k_2)/{\mathcal{D}} \, , \quad d_3 = 2i\hbar\a(b_0 - 3\ell_2)/{\mathcal{D}} \nonumber
 \ea
with\footnote{Incidentally, the constants $d_2$ and $d_3$ should
not be confused with $e_2\lrcorner d$ and $e_3\lrcorner d$,
repectively.}
 \ba
   \mathcal{D}= 4 \left[ a_0(2a_0 - 5k_1 + 3k_2 - 12k_3) + 2k_1(k_1 -3k_2 + 3k_3) - 18k_2k_3 + 2b_0^2 - 18\ell_2^2 \right] \, . \nonumber
 \ea
This result has been checked by the computer algebra system Reduce
and its package Excalc \cite{Hearn1093},\cite{Schrufer}. For the
Bjorken-Drell conventions $S^\a \rightarrow - S^\a$ and $W^\a
\rightarrow -i W^\a$. Now, by calculating $\mathcal{V}=6\kappa
(c_2 S + d_2 W)$ and $\mathcal{A}=6\kappa (c_3 S + d_3 W)$, we
obtain the irreducible parts of the torsion as
 \ba
     {}^{(1)}T^\a =0, \;\; {}^{(2)}T^\a =2\kappa  \vt^\a \w (c_2 S + d_2 W), \;\; {}^{(3)}T^\a =2\kappa e^\a \lrcorner {}^\star (c_3 S + d_3 W) \, .
 \ea
Therefore we can drop $a_1$ (or $k_1$) without loss of generality.
So, we will continue with non-vanishing coupling constants
$\alpha,a_0,\lambda_0,b_0,k_2=a_2/3,k_3=a_3/3,\ell_2=-\sigma_2/3$.
Thus,
 \ba
   & &  c_2=-\hbar(b_0 + 3\ell_2) / \mathfrak{D}  \quad \quad \; \; {\text{(pseudoscalar)}}\, , \label{eq:c2} \\
   & &  d_2= i\hbar \a(a_0 - 6k_3) / \mathfrak{D}  \quad \quad \, {\text{(scalar)}}\, , \label{eq:d2} \\
   & &  c_3= \hbar (2a_0 + 3k_2) / \mathfrak{D}  \quad \quad \; \; \; {\text{(scalar)}}\, , \label{eq:c3} \\
   & &  d_3= 2i\hbar \a(b_0 - 3\ell_2) / \mathfrak{D} \quad \quad {\text{(pseudoscalar)}}  \, , \label{eq:d3}
 \ea
with
 \ba
     \mathfrak{D} = 4\left[ a_0(2a_0 + 3k_2 - 12k_3) - 18k_2k_3 + 2b_0^2 - 18\ell_2^2 \right] \, .
 \ea

Here we recall that the coefficients $a_0,\lambda_0,k_1,k_2,k_3$
are scalars (since they contain an odd number of Hodge stars) and
the remaining ones $b_0,\ell_1,\ell_2$ are pseudoscalars
(containing an even number of Hodge stars). Because the
denominator contains squares of $b_0$ and $\ell_2$, it is a scalar
quantity. Thus, as $c_2$ and $d_3$ are pseudoscalars, $d_2$ and
$c_3$ are scalars. If we put the pseudoscalar coefficients to
zero, $b_0=\ell_2=0$, then we are left only with the scalars $d_2$
and $c_3$ as a consistent result. We also remark that for a viable
theory the constants have to be constrained by studying the
propagating modes (see Chen {\it et al} \cite{chen2009}). It is
clear that a vanishing  $\frak{D}$ would disqualify the model from
being physical. In other words, a closer investigation has to get
information also on the behavior of $\frak{D}$.

The next step is to rewrite the Dirac equation (\ref{diraceqn2}) in terms of Riemannian quantities.
The first job is to calculate the contortion via (\ref{contort})
 \ba
    K_{\a\b} = \kappa [2c_2 (S_\b \vt_\a - S_\a \vt_\b ) +2d_2 (W_\b \vt_\a - W_\a \vt_\b ) - c_3 {}^\star (\vt_{\a\b} \w S) - d_3 {}^\star (\vt_{\a\b} \w W)] \, .
 \ea
Then the Dirac equation (\ref{diraceqn2}) turns out to be
 \ba
 {}^\star \gamma \w \widetilde{D}\Psi + \left\{\frac{mc}{\hbar}
 - \frac{3\kappa}{4} \left[ S_\a \gamma^\a(c_3\gamma_5- 4i\a c_2) + W_\a \gamma^\a(d_3\gamma_5- 4i\a d_2) \right] \right\} \Psi \, \eta=0 \,
 . \label{diraceqnrieman}
 \ea
We obtained the parity-conserving terms $\sim c_3 S_\a \gamma^\a
\gamma_5 \Psi$ and $\sim 4i\a d_2 W_\a \gamma^\a \Psi$, and the
parity-violating term $\sim \gamma^\a(-4i\a c_2 S_\a + d_3
\gamma_5 W_\a)\Psi$. But for $\a =0$, the parity-violating term
and the second one of the parity conserving terms drop out and we
are left with only the first parity-conserving term.

Finally we split the FIRST equation into a Riemann part and a torsion part.
Let us first decompose the curvature 2-form as Riemann 2-form plus contortion terms
 \ba
    R_\a{}^\b &=& d(\widetilde{\Gamma}_\a{}^\b -K_\a{}^\b) - (\widetilde{\Gamma}_\a{}^\g -K_\a{}^\g)\w (\widetilde{\Gamma}_\g{}^\b -K_\g{}^\b) \nonumber \\
        &=& \widetilde{R}_\a{}^\b - \widetilde{D} K_\a{}^\b -K_\a{}^\g \w K_\g{}^\b \, .
 \ea
Then, with the Einstein 3-form $\widetilde{G}_\a = \frac{1}{2}\widetilde{R}^{\b\g} \w \eta_{\a\b\g}$,
we can write the FIRST equation (\ref{firsteqn}) in terms of Riemannian quantities plus Dirac field
 \ba
    a_0\widetilde{G}_\a + \lambda_0 \eta_\a =\kappa [ \widetilde{\Sigma}_\a - \frac{\hbar}{4} (\widetilde{D} S^\b) \w \vt_{\a\b}
    - \frac{i \a \hbar}{2}  (\widetilde{D} W^\b) \w \eta_{\a\b} ]  \nonumber \\
    +\kappa^2 [ C_S S_\b S^\b + C_W W_\b W^\b +  C_{SW} S_\b W^\b] \eta_\a  \label{firsteqnrieman}
 \ea
where
 \ba
    C_S &=& -a_0(4c_2^2 - c_3^2)  + 30k_2c_2^2 - 6k_3c_3^2 - \hbar c_3 - 4c_2c_3(b_0 - 9\ell_2) \, , \label{eq:B} \\
    C_W &=& -a_0(4d_2^2 - d_3^2)  + 30k_2d_2^2 - 6k_3d_3^2 + 2i \a \hbar d_2 - 4d_2d_3(b_0 + 9\ell_2) \, , \label{eq:cw} \\
    C_{SW} &=& -2a_0(4c_2d_2 - c_3d_3)  + 60k_2c_2d_2 - 12k_3c_3d_3 - \hbar d_3 + 2i \a \hbar c_2 \nonumber \\
            & &  \quad \quad \quad - 4(d_2c_3+c_2d_3)(b_0 - 9\ell_2) \, .  \label{eq:csw}
 \ea
Whereas $C_S,C_W$ are scalars, $C_{SW}$ is a pseudo scalar. In
(\ref{firsteqnrieman}) we decomposed the Dirac energy-momentum
current (\ref{direnmomcurrent}) according to
 \ba
  \Sigma_\a = \widetilde{\Sigma}_\a - \frac{\kappa \hbar}{2} \left\{ c_3 \left( 2S_\b S^\b \eta_\a + S_\a S_\b \eta^\b \right) + d_3 \left( 2S_\b W^\b \eta_\a + S_\a W_\b \eta^\b \right) + 2 d_2 S^\b W^\g \vt_{\a\b\g} \right. \nonumber \\
   \left. - i\a \left[ 2d_2 \left( 2W_\b W^\b \eta_\a + W_\a W_\b \eta^\b \right) + 2c_2 \left( 2S_\b W^\b \eta_\a + W_\a S_\b \eta^\b \right) + c_3 S^\b W^\g \vt_{\a\b\g} \right] \right\} \, ,
 \ea
with
 \ba
    \widetilde{\Sigma}_\a =\frac{i\hbar}{2} [ (1-i\a) \overline{\Psi} \gamma_\b (\widetilde{D}_\a \Psi) - (1+i\a) (\widetilde{D}_\a \overline{\Psi}) \gamma_\b \Psi]\eta^\b \, . \label{eq:emDiracRiem}
 \ea

Now we note that the Riemannian field equations
(\ref{diraceqnrieman}) and (\ref{firsteqnrieman}) are derived from
the (hermitian) lagrangian 4-form $L_{\text{tot}}= V_{EH\lambda} +
L_D + L_{\text{int}} + \Lambda_\a \w T^\a$ where the lagrange
multiplier $\Lambda_\a$ is a 2-form, $L_D$ is the non-minimally
coupled the Dirac lagrangian given by (\ref{10-4.10}),
$V_{EH\lambda}$ is the Einstein-Hilbert lagrangian with the
cosmological constant
 \ba
    V_{EH\lambda}&=&-\frac{a_0}{2\kappa} R^{\a\b} \w \eta_{\a\b} - \frac{\lambda_0}{\kappa} \eta  \label{eq:EHlam}
 \ea
and $L_{\text{int}}$ is the interaction lagrangian
 \ba
    L_{\text{int}} &=& \kappa (E_S S_\a S^\a +  E_W W_\a W^\a +  E_{SW} S_\a W^\a) \eta \, ,
 \ea
with the coefficients
 \ba
    E_S &=& -3a_0(4c_2^2 -c_3^2) - 12 b_0 c_2c_3 + 18 k_2 c_2^2 - 18k_3c_3^2 + 36 \ell_2 c_2c_3 - 3 \hbar c_3/2 \, , \\
    E_W &=& -3a_0(4d_2^2 -d_3^2) - 12 b_0 d_2d_3 + 18 k_2 d_2^2 - 18k_3d_3^2 + 36 \ell_2 d_2d_3 + 3i \alpha \hbar d_2 \, , \\
    E_{SW} &=& -6a_0(4c_2d_2 -c_3d_3) - 12 b_0 (c_2d_3 + d_2c_3) + 36 k_2 c_2 d_2 - 36k_3c_3 d_3 \nonumber \\
             & & \quad  \quad + 36 \ell_2 (c_2d_3 + d_2c_3) - 3 \hbar d_3/2 + 3i \alpha \hbar c_2\, .
 \ea

 \subsection*{The Einstein-Dirac Theory with $\lambda_0$}

For comparison with the literature, we want to summarize briefly
the minimally coupled Einstein-Dirac theory with cosmological
constant. It is defined by the total lagrangian 4-form
$L_{\text{tot}}= V_{EH\lambda} + L_D + \Lambda_\a \w T^\a $ where
$V_{EH\lambda}$ is given by (\ref{eq:EHlam}), $L_D$ is the
minimally coupled Dirac lagrangian that is given by
(\ref{10-4.10}) with $\alpha =0$ and $\Lambda_\a$ is the lagrange
multiplier 2-form whose variation yields $T^\a =0$. The
excitations are calculated as
 \ba
     H_\a = 0 \, , \quad H_{\a\b} = \frac{a_0}{2\kappa} \eta_{\a\b} \, .
 \ea
Then the gauge currents are obtained as
 \ba
     E_\a = - \frac{a_0}{2\kappa} R^{\b\g} \w \eta_{\a\b\g} - \frac{\lambda_0}{\kappa} \eta_\a \, , \quad
     E_{\a\b} = 0\, .
 \ea
Since the torsion is zero, all the concerned Riemannian quantities
will be marked by a tilde. Now we write down the modified FIRST
equation, $DH_\a -E_\a -D\Lambda_\a = \Sigma_{\a}$, the modified
SECOND equation, $DH_{\a\b} - E_{\a\b} - \frac{1}{2} (\vt_\a \w
\Lambda_\b -  \vt_\b \w \Lambda_\a) = \tau_{\a\b}$ and the Dirac
equation, respectively
 \ba
      \frac{a_0}{2\kappa} \widetilde{R}^{\b\g} \w \eta_{\a\b\g} + \frac{\lambda_0}{\kappa} \eta_\a -\widetilde{D}\Lambda_\a &=& \widetilde{\Sigma}_\a \, , \label{eq:modFIRST}\\
     \frac{a_0}{2\kappa} \widetilde{D}\eta_{\a\b} - \frac{1}{2}(\vt_\a \w \Lambda_\b - \vt_\b \w \Lambda_\a) &=& \frac{\hbar}{4}S^\g \vt_{\a\b\g} \, , \label{eq:modSECOND}\\
     {}^\star \gamma \w \widetilde{D}\Psi + \frac{mc}{\hbar} \Psi \, \eta &=&0 \, , \label{diraceqnRiem}
 \ea
where the Riemannian Dirac energy-momentum current
$\widetilde{\Sigma}_\a$ is the same as (\ref{eq:emDiracRiem}) with
$\a =0$. By using the identity $\widetilde{D}\eta_{\a\b}=0$ we
calculate $\Lambda_\a= -\frac{\hbar}{4}S^\b \vt_{\a\b}$ from
(\ref{eq:modSECOND}). The substitution of this into
(\ref{eq:modFIRST}) gives rise to the following
 \ba
  a_0\widetilde{G}_\a + \lambda_0 \eta_\a =\kappa \widetilde{\Sigma}_\a - \frac{\kappa \hbar}{4} (\widetilde{D}S^\b) \w \vt_{\a\b} \, . \label{firsteqnRiem}
 \ea

\section{Concluding Remarks}

In this work we considered the Dirac field non-minimally coupled
to the gravitational field in the framework of the Poincare gauge
theory. The gravitational lagrangian contains terms at most linear
in curvature and quadratic in torsion, both  of odd and even
parity. This is the most general $R+T^2$-type lagrangian with
eight parameters, $\alpha,\lambda_0,a_0,b_0,k_1,k_2,k_3,\ell_2$,
all together. The field equation, obtained by varying the total
lagrangian with respect to the Lorentz connection, was solved with
respect to torsion in terms of the Dirac vector and the Dirac
axial vector (\ref{eq:tors}). Our first result is that as the
tensor piece of the torsion, ${}^{(1)}T^\a$, is zero and the
vector part, ${}^{(2)}T^\a$, and the axial part, ${}^{(3)}T^\a$,
are nonzero. Thus we set $k_1=0$ without loss of generality and
the number of the parameters of our model decreased to seven.
However, this is still the most general $R+T^2$ model in the
literature. Then by using the decomposition of the Lorentz
connection in terms of its Riemannian piece and contortion, the
Dirac equation is decomposed into a Riemannian piece and torsional
terms, see eqn(\ref{diraceqnrieman}). Here we observe that the
parity-violating term $\sim \gamma^\a(S_\a \pm \gamma_5 W_\a)\Psi$
is proportional to non-minimal coupling parameter, $\alpha$. After
that we applied the same strategy to the field equation obtained
by variation with respect to the orthonormal coframe, see
eqn(\ref{firsteqnrieman}). Consequently, the FIRST equation turned
out to be the form of the familiar Einstein-Dirac equation plus
correction terms. Then we obtained a Riemannian lagrangian 4-form
equivalent to our Poincare gauge theory of gravity. Finally we
summarized the minimally coupled Einstein-Dirac theory with
$\lambda_0$ for easy comparison of our results with the
literature. Accordingly when $\alpha=0$ we realized that the
relevant quantities are $c_3$ given by (\ref{eq:c3}) and $C_S$
given by (\ref{eq:B}). Interestingly $c_2$ given by (\ref{eq:c2})
does not appear explicitly in the Dirac equation
(\ref{diraceqnrieman}) in this case. Thus the case of that
$\alpha=0$, $c_3=0$ and $C_S=0$, including the possibility $k_2 =
-2a_0/3$ and $\ell_2 = -b_0/3$ as well, corresponds to the
minimally coupled Einstein-Dirac theory. For the minimally coupled
Einstein-Cartan-Dirac theory with cosmological constant
($\a=0,k_2=k_3=b_0=\ell_2=0$) the relevant coefficients turn out
to be $C_S=-3\hbar^2/16a_0$ and $c_3=\hbar/4a_0$. In the minimally
coupled weak Poincare gauge theory with even parity
($\alpha=b_0=\ell_2=0$) they are nonzero. In the opposite case
with odd parity ($\a=a_0=k_2=k_3=0$) both $c_3$ and $C_S$ vanish.
For our case they shift compared with the literature.

In Ref. \cite{lammerzahl1997} the authors reanalyzed the
Hughes-Drever type experiments carried out for testing the
anisotropy of mass and anomalous spin couplings and deduced a
constraint on the axial torsion by $\left| S_\alpha \right| \leq
10^{-15} \, \mbox{m}^{-1}$. Since they considered minimal Dirac
couplings ($\alpha =0$) it was not possible to test other parts of
the torsion tensor. Then they argue that one needs higher spin
equations for a coupling to the trace and the traceless part of
torsion tensor. Thus to repeat a similar analysis for our Dirac
equation (\ref{diraceqnrieman}) may give insights on the
constraints of vector components of torsion. A decade later
Kostelecky {\it et al} exploited newer experimental searches
(Zeeman measurements with a dual maser and studies of a
spin-polarized torsion pendulum)  for Lorentz violation in order
to extract new constraints on torsion components down to levels of
order $10^{-15} \, \mbox{m}^{-1}$ \cite{kostelecky2008}. Their
analysis were performed through torsion nonminimal couplings to
standard-model fields. Some combinations of our parameters
correspond to their parameters $\xi_1^{(4)}, \xi_2^{(4)},
\xi_3^{(4)}, \xi_4^{(4)}$. In another work the authors discussed
the possibility to perform and use the exact Foldy-Wouthuysen
transformation for the Dirac spinor coupled to different CPT and
Lorentz violating terms \cite{goncalves2009}. In accord with their
result there may be a mixing between the magnetic and torsion
fields, and the strongness of magnetic field may compensate the
weakness of torsion. Consequently, the mixture term may affect the
motion of a test particle in a notable way. When our work is
compared with that of \cite{goncalves2009}, it is realized that
their four parameters $ a_\mu , b_\mu , c^{\mu \nu}, d^{\mu \nu}$
are some combinations of ours.

\section*{Acknowledgments}

The author thanks F.W. Hehl for reading the paper and for stimulating comments.

\appendix

\section{Irreducible Decomposition of Torsion}

Torsion is decomposed as
 \ba
   T^\alpha = {}^{(1)}T^\alpha + {}^{(2)}T^\alpha + {}^{(3)}T^\alpha
 \ea
in terms of
 \ba
    & & {}^{(2)}T^\alpha = - \frac{1}{3} \mathcal{V} \wedge \vartheta^\alpha \, ,\\
    & & {}^{(3)}T^\alpha =  \frac{1}{3}   {}^\star(\mathcal{A} \w \vartheta^\a) \, ,\\
    & & {}^{(1)}T^\alpha = T^\alpha - {}^{(2)}T^\alpha - {}^{(3)}T^\alpha \, ,
 \ea
where $\mathcal{V}= e_\a \lrcorner T^\a$ and $\mathcal{A}= {}^\star(\vt_\a \w T^\a)$.
The irreducible components of torsion satisfy
 \ba
      {}^{(1)}T^\alpha \w \vartheta_\a =0 \; , \;\; {}^{(2)}T^\alpha \w \vartheta_\a =0 \; , \;\;
     e_\a \lrcorner {}^{(1)}T^\alpha = 0 \; ,\;\; e_\a \lrcorner {}^{(3)}T^\alpha = 0 \, .
 \ea
Thus the ${}^{(I)}T^\alpha$'s are orthogonal in the following two senses:
 \ba
  & &  {}^{(I)}T^\alpha \wedge {}^\star {}^{(J)}T_\alpha ={}^{(J)}T^\alpha \wedge {}^\star {}^{(I)}T_\alpha \left\{ \begin{array}{cl}
                                                     \neq 0 & \mbox{if} \;\;I=J \\
                                                     =0 & \mbox{if} \;\;I \neq J
                                                   \end{array} \right.\\
   & &  {}^{(I)}T^\alpha \wedge {}^{(J)}T_\alpha={}^{(J)}T^\alpha \wedge {}^{(I)}T_\alpha \left\{ \begin{array}{cl}
                                                     \neq 0 & \mbox{if} \; \; I = J=1 \\
                                                     \neq 0 & \mbox{if} \; \; I=2, \; J=3\\
                                                     =0     & {\text{otherwise}}\\
                                                   \end{array} \right.
 \ea

\section{Irreducible Decomposition of Energy-momentum 3-form in 4 Dimensions}

Any vector-valued 3-form $\Sigma_\a$ in 4 dimensions can be decomposed as
 \ba
   \Sigma_\a = {}^{(1)}\Sigma_\a + {}^{(2)}\Sigma_\a + {}^{(3)}\Sigma_\a
 \ea
in terms of
 \ba
    & & {}^{(2)}\Sigma_\a = \frac{1}{2} \mathcal{F} \w \vt_\a \, ,\\
    & & {}^{(3)}\Sigma_\a = -\frac{1}{4} e_\a \lrcorner \mathcal{G} \, ,\\
    & & {}^{(1)}\Sigma_\a = \Sigma_\a - {}^{(2)}\Sigma_\a - {}^{(3)}\Sigma_\a \, ,
 \ea
where $\mathcal{F} = e^\a \lrcorner \Sigma_\a$ and $\mathcal{G}= \Sigma_\a \w \vt^\a$. The irreducible parts satisfy
 \ba
    {}^{(1)}\Sigma^\a \w \vt_\a=0 \; , \;\;  {}^{(2)}\Sigma^\a \w \vt_\a=0  \; , \;\; e_\a \lrcorner {}^{(1)}\Sigma^\a =0
     \; , \;\;    e_\a \lrcorner {}^{(3)}\Sigma^\a =0 \, .
 \ea

\section{Irreducible Decomposition of Spin 3-form in 4 Dimensions}

Any bivector-valued 3-form $\tau_{\a\b}=-\tau_{\b\a}$ in 4 dimensions can be decomposed as
 \ba
   \tau_{\a\b} = {}^{(1)}\tau_{\a\b} + {}^{(2)}\tau_{\a\b} + {}^{(3)}\tau_{\a\b}
 \ea
in terms of
 \ba
    & & {}^{(2)}\tau_{\a\b} = \frac{1}{3} (e_\a \lrcorner \mathcal{W}_\b -e_\b \lrcorner \mathcal{W}_\a) \, ,\\
    & & {}^{(3)}\tau_{\a\b} =  -\frac{1}{6} \mathcal{B} \w \vt_{\a\b} \, ,\\
    & & {}^{(1)}\tau_{\a\b} = \tau_{\a\b} - {}^{(2)}\tau_{\a\b} - {}^{(3)}\tau_{\a\b} \, ,
 \ea
where $ \mathcal{W}_\a = \tau_{\a\b} \w \vt^\b$ and
$\mathcal{B}=e_\a \lrcorner e_\b \lrcorner \tau^{\a\b}$. The
irreducible parts satisfy
 \ba
     e_\a \lrcorner e_\b \lrcorner {}^{(1)}\tau^{\a\b} =0 \; , \;\;  e_\a \lrcorner e_\b \lrcorner {}^{(2)}\tau^{\a\b} =0 \; , \;\;
     {}^{(1)}\tau^{\a\b} \w \vt_\b=0  \; , \;\; {}^{(3)}\tau^{\a\b} \w \vt_\b=0 \, .
 \ea
For more information on the decompositions of the related
quantities one can consult \cite{Baekler:2010fr}.

\end{document}